\documentclass[
 preprint,
 amsmath,amssymb,
 aps,
 prd
]{revtex4-2}

\usepackage{amsmath}
\usepackage{amssymb}
\usepackage{graphicx}
\usepackage{hyperref}

\newcommand{\recip}[1]{{\frac{1}{#1}}}
\newcommand{\trecip}[1]{{\tfrac{1}{#1}}}
\begin{document}

\title{Perturbative solutions for the bumblebee field in a Schwarzschild background}

\author{Demetre Seturidze}
\email{demetre.seturidze@tufts.edu}
 \altaffiliation[\\Present address: ]{Department of Physics, Tufts University}
\author{Don Colladay}
 \email{colladay@ncf.edu}
\affiliation{
Department of Physics, New College of Florida
}

\date{\today}

\begin{abstract}
In this paper, we investigate the behavior of a vector field subject to a Lorentz-violating quartic potential on a fixed Schwarzschild background. We analyze the equilibrium configuration of the field and its linear perturbations. We show that the equilibrium solution and spherically symmetric perturbations are governed by a quadratic function whose properties determine their qualitative behavior. By requiring the background field to be real and finite, we derive bounds on the allowed parameters and identify critical cases in which the monopole perturbations become enhanced. We further derive asymptotic expressions for multipole perturbations near the event horizon and at spatial infinity. 
\end{abstract}

\maketitle

\section{Introduction}

    In 1989, Kosteleck\'y and Samuel \cite{kst_sml:1989} showed that Lorentz symmetry could be spontaneously broken in certain low-energy limits of string theory. This discovery motivated significant efforts to characterize violations of local Lorentz invariance, most notably with the Standard Model Extension \cite{kst_cld:1997, kst_cld:1998}, and to impose observational constraints on symmetry-breaking parameters \cite{kst_rsl:2011}.

    Among the simplest and most extensively studied Lorentz-violating models is the bumblebee model, in which a vector field acquires a nonzero vacuum expectation value through a constraint or a symmetry-breaking potential. Due to its simplicity, the bumblebee model has served as a useful framework for investigating spontaneous Lorentz symmetry breaking in a broad range of contexts, including modified gravity, black-hole spacetimes, cosmology, and matter dynamics \cite{bluhm:2007, bluhm:2008, capelo:2015, casana:2018, delhom:2022, ding:2020, jha:2021, kst:2004, maluf:2014, nilsson:2026}. 
    
    Much of the existing literature has focused on constructing consistent background solutions and analyzing their consequences for modified gravitational dynamics or particle motion. Comparatively less attention has been devoted to the perturbative dynamics of the bumblebee field itself on fixed curved spacetimes. Recent works have obtained exact static solutions in a variety of geometries \cite{li:2026, liu:2026, xu:2023, xu:2026, zhu:2026}, and linearized perturbations around the vacuum expectation value have been studied in flat spacetime \cite{bailey:2025}. A detailed treatment of such perturbations on curved backgrounds remains largely unexplored.

    In this paper, we conduct a linear perturbative analysis of the equations of motion of the bumblebee field in a fixed Schwarzschild spacetime. We employ the quartic potential
    \begin{equation}
    V\!\left(B^2\right) = \trecip2 \kappa \big(B^2-b^2\big)^2,
    \end{equation}
    because it dynamically realizes the symmetry-breaking vacuum without imposing a constraint through a Lagrange multiplier. More generally, the analysis applies to any symmetry-breaking potential whose expansion about the vacuum is quadratic to leading order.
    
    The system admits a minimizing background configuration \(b_\mu\). Although we do not impose a Lagrange-multiplier constraint on the full field, the background configuration necessarily satisfies the norm condition
    \begin{equation}
        b_\mu b^\mu \equiv b^2 = \gamma, \label{eq:constr}
    \end{equation}
    where \(\gamma\) is a constant.
    Small deviations from this configuration are introduced through the expansion
    \begin{equation}
        B_\mu = b_\mu + \varepsilon f_\mu,
    \end{equation}
    where \(\varepsilon\) is a perturbative parameter. Expanding the field equations to first order in \(\varepsilon\) generates a linear system about the background configuration. 
    
    At zeroth order, the background configuration satisfies the vacuum field equations together with the norm condition \eqref{eq:constr}. At first order, the perturbation \(f_\mu\) satisfies a linear equation whose source is proportional to its projection onto the background field, \(f_\mu b^\mu\). We construct static, spherically symmetric background configurations and then study the first-order perturbations. In the monopole sector, we obtain explicit time-dependent solutions, while in the higher multipole sectors, we derive asymptotic, time-independent solutions near the horizon and at spatial infinity.

\section{Perturbative Expansion}
    \label{sec:pert}
    
    We assume a fixed Schwarzschild background in the Gullstrand--Painlev\'e coordinate system \cite{gullstrand:1922, painleve:1921} with the metric
    \begin{equation}
        \begin{gathered}
        ds^2 = U\!(r)dt^2-2\sqrt{\frac{s}r} dtdr -dr^2-r^2d\Omega^2, \\ U(r)=1-\frac{s}r,
        \end{gathered}
    \end{equation}
    where \(s=2GM\) is the Schwarzschild radius. The metric components in this coordinate system remain finite at the event horizon \(r = s\), allowing us to enforce physicality conditions on the field components across the event horizon. It is related to the Schwarzschild coordinate system by the following transformation
    \begin{equation}
        dt_{GP} = dt_{S} + \frac{\sqrt{\frac sr}}U dr.
    \end{equation}
    
    We consider the bumblebee field as a test field propagating on this fixed Schwarzschild background. The metric is therefore nondynamical, and we include no Einstein--Hilbert term in the action below. Furthermore, the Ricci tensor identically vanishes on a Schwarzschild background, so terms involving explicit couplings to the Ricci tensor also vanish. We therefore take the action of the bumblebee field to be the following:
    \begin{equation}
        S=\int dx^4 \sqrt{-g} \left(-\recip4B_{\mu\nu} B^{\mu\nu}-\recip2 \kappa \bigl(B^2 -b^2\bigr)^2\right),
    \end{equation}
    where \(B^2 = B_\mu B^\mu\) and \(b^2 = b_\mu b^\mu\). 
    
    The bumblebee action resembles that of Proca theory in that both include a Maxwell-like kinetic term \(B_{\mu\nu} B^{\mu\nu}\), as well as a potential term. Unlike the Proca potential, the bumblebee potential is quartic and minimized at a nonzero field norm
    \begin{equation}
        B^2=b^2,
    \end{equation}
    allowing the vector field to acquire a nonzero vacuum expectation value. The corresponding equations of motion are
    \begin{equation}
        \nabla_\mu B^{\mu\nu} = 2\kappa\bigl( B^2 - b^2 \bigr) B^\nu.
        \label{eq:setup}
    \end{equation}

    We write the field components \(B_\mu\) in terms of the background configuration \(b_\mu\) and a perturbation \(f_\mu\) parameterized by the variable \(\varepsilon\). Expanding \eqref{eq:setup} to first order in \(\varepsilon\) gives
    \begin{equation}
    {
    \begin{gathered}
    B_\mu \:=\: b_\mu + \varepsilon f_\mu,\\[1.5ex]
        \nabla_\mu b^{\mu\nu} = 0, \qquad \nabla_\mu f^{\mu\nu} = 4\kappa (f_\mu b^\mu) b^\nu,
        \label{eq:motion}
    \end{gathered}
    }
    \end{equation}
    where \(b_\mu\) satisfies the norm condition \eqref{eq:constr}, and \(b^{\mu\nu}\) and \(f^{\mu\nu}\) are the components of the antisymmetric field tensors
    \begin{equation}
        \begin{gathered}
        b^{\mu\nu} = \nabla^\mu b^\nu - \nabla^\nu b^\mu, \\ f^{\mu\nu} = \nabla^\mu f^\nu - \nabla^\nu f^\mu.
        \end{gathered}
    \end{equation}

    \section{The Equilibrium Configuration}
    Taking the equilibrium solution \(b_\mu\) to be time-independent and rotationally invariant gives
    \begin{equation}
    \recip{r^2}\partial_{r} \,r^2 \partial_{r} \, b_{t} = 0 \quad\implies\quad b_{t} = \alpha + \beta U.
    \end{equation}
    Here, \(\beta\) parameterizes the strength of the unperturbed antisymmetric field tensor, while \(\alpha\) is an integration constant.
    The condition \eqref{eq:constr} produces a quadratic equation for the radial component \(b_{r}\)
    \begin{equation}
    \begin{aligned}
    &\gamma = b_\mu b^\mu = b_{t}^2 -2\sqrt{\frac sr} \, b_{r} \,b_{t} -Ub_{r}^2 \\\implies &b_{r} = \recip U \left[\pm\sqrt{b_{t}^2 - \gamma U} - b_{t} \sqrt{\frac sr}\right].
    \end{aligned}
    \end{equation}
    
    The quantity \(b_{t}^2 - \gamma U\) is of central importance throughout this paper. We will refer to it as the \textbf{characteristic quadratic} of the solution system
    \begin{equation}
        C_{\alpha\beta\gamma}(U) = \alpha^2 + (2\alpha\beta- 
        \gamma)U + \beta^2U^2,
    \label{eq:char_quad}
    \end{equation}
    since it governs both the existence of solutions and the behavior of the perturbations.
    
    In the \(\beta=0\) case, this object collapses to a first-order polynomial. Otherwise, it has roots at \(U_0\pm i\Delta U\), where
    \begin{equation}
        U_0 =\frac{\gamma-2\alpha\beta}{2\beta^2}, \quad\Delta U = \frac{\sqrt{\gamma(4\alpha\beta -\gamma)}}{2\beta^2}.
    \end{equation} 

    We require the field components to remain finite throughout the region \(r > 0\). In particular, we assert that the components remain finite across the event horizon. The Taylor expansion of \(Ub_{ r}\) near the event horizon is
    \begin{multline}
        Ub_r \approx \big(\pm|\alpha| - \alpha\big) \\+ \recip2\bigg[-2\beta + \alpha \pm\left(\text{sign}(\alpha) 2\beta -\frac{\gamma}{|\alpha|}\right)\bigg] U.
        \label{eq:Taylor_Ub_r}
    \end{multline}
    The constant term in \eqref{eq:Taylor_Ub_r} must vanish in order for \(b_{ r}\) to remain finite, which fixes the sign choice \(\pm\) to match the sign of \(\alpha\). This leaves us with
    \begin{equation}
        Ub_r \approx \recip2 \left(\alpha - \frac{\gamma}{\alpha}\right) U \quad\implies\quad b_r \approx \recip2 \left(\alpha - \frac{\gamma}{\alpha}\right),
    \end{equation}
    whereby regularity of \(b_{ r}\) imposes \(\alpha \ne 0\).
    
    Therefore, we may express the complete equilibrium solution as follows
    \begin{equation}
    {
    \begin{gathered}
    b_{ t} = \alpha + \beta U, \\ b_{ r} = \recip{U}\left(\text{sign}(\alpha)\,\sqrt{C_{\alpha\beta\gamma}(U)} \; - (\alpha + \beta U ) \sqrt{\frac sr}\right), \\\alpha \ne 0.
    \end{gathered}
    }
    \label{eq:quadratic_base_soln}
    \end{equation}

    Requiring the field to be real for \(r>0\) gives
    \begin{equation}
    \min_{U \in (-\infty, 1]} C_{\alpha\beta\gamma}(U) \ge 0.
    \label{eq:b_realness}
    \end{equation}
    
    If \(|\beta| \ge |\alpha|\), then \eqref{eq:b_realness} corresponds to the region \(I\) in the parameter space
    \begin{equation}
        \gamma \in I=\big[\!\min(0, 4\alpha\beta), \, \max(0, 4\alpha\beta)\big], \quad \text{for } |\beta| \ge |\alpha|.
        \label{eq:range}
    \end{equation}
    For any combination of parameters in the interior of this region, \(\Delta U\) is real and \(C_{\alpha\beta\gamma}\) is positive for all \(U\). The boundaries of \(I\) correspond to a degeneracy in which \(\Delta U\) vanishes and the characteristic quadratic attains a double root at \(U=U_0\).
    \begin{equation}
        \begin{aligned}
        \gamma = 0 &\:\implies\: U_0=\frac{-\alpha}{\beta}, \\
        \gamma = 4\alpha\beta &\:\implies\: U_0= \frac{\alpha}{\beta}.
        \end{aligned}
        \label{eq:bounds}
    \end{equation}
    In these cases, \(\sqrt{C_{\alpha\beta\gamma}}\) is not differentiable near \(U=U_0\) and the radial component \(b_r\) has a cusp. We refer to these parameter choices as \textbf{critical configurations}.

    If \(|\beta| < |\alpha|\), then the larger value of \(U_0\) given in \eqref{eq:bounds} falls outside the desired region \(U \in (-\infty, 1]\), and the parameter space can be extended until the smaller root of \(C_{\alpha\beta\gamma}\) occurs at \(U=1\). The complete admissible region of the parameter space in this case is
    \begin{equation}
        \gamma \in I^*= \big[\! \min(0, 4\alpha\beta), \, (\alpha+\beta)^2\big], \quad \text{for }|\beta|<|\alpha|.
        \label{eq:ext_range}
    \end{equation}
    The upper limit \(\gamma = (\alpha+\beta)^2\) now corresponds to a cusp at \(U=1\), equivalent to \(r=\infty\).

    \begin{figure}[tbp]
        \includegraphics{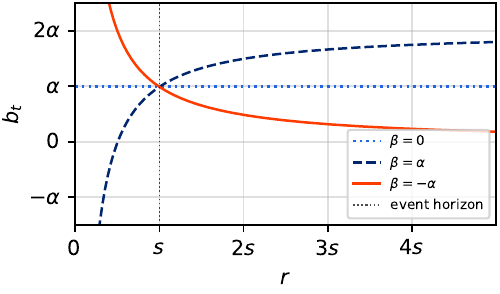}
        \caption[Temporal component of the equilibrium solution]{Behavior of \(b_{ t}(r)\) for selected parameter choices.}
        \label{fig:b_t}
    \end{figure}

    \begin{figure}[tbp]
        \includegraphics{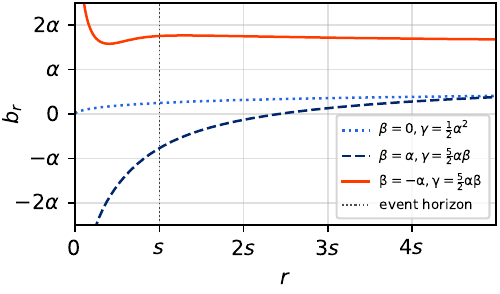}
        \caption[Radial component of the equilibrium solution, non-critical]{Behavior of \(b_{ r}(r)\) for noncritical cases.}
        \label{fig:b_r}
    \end{figure}

    \begin{figure}[tbp]
        \includegraphics{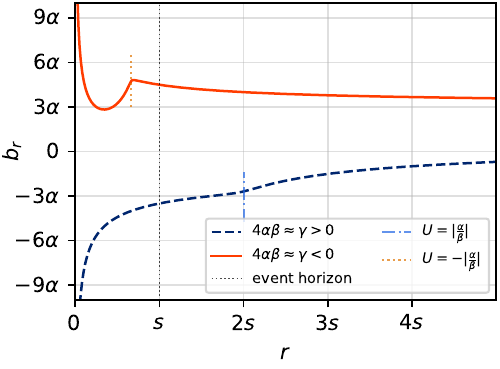}
        \caption[Radial component of the equilibrium solution, critical]{Behavior of \(b_{ r}(r)\) for critical cases with $|\beta| = 2|\alpha|$.}
        \label{fig:b_r_critical}
    \end{figure} 

    The requirement that \(b_\mu\) be differentiable for \(r>0\) excludes the boundaries of \eqref{eq:range} and \eqref{eq:ext_range} from the admissible parameter space, but does not prevent near-critical behavior in configurations arbitrarily close to these boundaries. As we show in the following section, the first-order perturbations for such configurations become increasingly large and vary rapidly near the roots of \(C_{\alpha\beta\gamma}\) and diverge in the critical limit. This behavior suggests that the linearized theory breaks down near the boundaries of the parameter space. 
    
    \section{Perturbations}

    Perturbing the field about the background configuration \(b_\mu\) produces
    \begin{equation}
    \begin{aligned}
        B_\mu&= b_\mu + \varepsilon f_\mu, \\[1.5ex]
        \nabla_\mu f^{\mu\nu} & =4\kappa (f_\mu b^\mu) b^\nu.
    \end{aligned}
    \label{eq:pert}
    \end{equation}
    We take \(\varepsilon\) to be small and assume that \(f_\mu\) vanishes at infinity.

    \subsection{Time-Dependent Monopoles}

    We assume spherical symmetry and apply a temporal Fourier transform to \(f_\mu\)
    \begin{equation}
        \hat f_\mu(\omega, r) = \int_{\mathbb R} \frac{dt}{\!\!\sqrt{2\pi}}\,e^{i\omega t} f_{\mu}(t, r),
    \end{equation}
    reducing the perturbed equations of motion \eqref{eq:pert} to the following pair
    \begin{align}
        \recip{r^2} \partial_r r^2 \left(i\omega \hat f_r + \partial_r \hat f_t \right)&=-4\kappa \left(\hat f_t b^t + \hat f_r b^r\right) b^t, 
        \label{i}\\
        -\omega^2  \hat f_{r} - i\omega \partial_{r}  \hat f_{t} &= -4\kappa\left(\hat f_tb^t + \hat f_rb^r\right) b^r.
        \label{ii}
    \end{align}
    The raised components \(b^\mu\) are given in \eqref{eq:raised_gp}.
    
    We define the auxiliary quantity \(\mu = r^2 \left(i\omega \hat f_{ r} + \partial_{r} \hat f_{ t}\right)\) representing the radial flux of the perturbation through a sphere of radius \(r\). Exploiting the similarity of the right-hand sides produces a first-order equation for \(\mu\)
    \begin{equation}
    \partial_{r}\mu - i\omega \frac{b^{ t}}{b^{ r}}  \mu \:=\: 0,
    \end{equation}
    which admits a general solution of the form
    \begin{equation}
    \mu(\omega, r)=  sa(\omega)e^{i\omega Z(r)}, \qquad Z(r)=\int dr \frac{b^{ t}}{b^{ r}},
    \label{eq:monopole_flux}
    \end{equation}
    where \(a(\omega)\) is an arbitrary complex-valued distribution in \(\omega\), and \(Z\) functions as the effective phase of radial oscillations. We introduce a factor of \(s\) for later convenience.
    
    By the definition of \(\mu\), we write the radial component as
    \begin{equation}
    \hat f_{ r} = \recip{i\omega} \left(\frac{\mu}{r^2} -\partial_{r} \hat f_{ t}\right).
    \label{d}
    \end{equation}
    Substituting this into \eqref{ii} produces 
    \begin{equation}
    \partial_{r} \,\hat f_{ t} - i\omega \frac{b^{ t}}{b^{ r}} \hat f_{ t} = a(\omega)e^{i\omega Z}\left(1 - \frac{\omega^2}{4\kappa C_{\alpha\beta\gamma}(U)} \right)\partial_r U.
    \end{equation}
    We apply the method of integrating factors and assert that the perturbations vanish at infinity. Together with \eqref{d}, this produces
    \begin{equation}
    {
    \begin{aligned}
        \hat f_{ t} &=  a(\omega)e^{i\omega Z} \left[\frac{\omega^2}{4\kappa} \big(F_{\alpha\beta\gamma}(1) - F_{\alpha\beta\gamma}(U)\big)-\frac{s}{r}\right],\\[1.5ex]
        \hat f_{ r} &=  - \frac{b^t}{b^r}\hat f_t -a(\omega)e^{i\omega Z} \! \left[\frac{i \omega \partial_r U}{4\kappa C_{\alpha\beta\gamma}(U)}\right].
        \end{aligned}
    } 
    \label{eq:quadratic_l0_soln}
    \end{equation}
    In the time-independent case \(\omega = 0\), the background field and perturbations are orthogonal. General monopole solutions can be constructed by applying an inverse Fourier transform to \eqref{eq:quadratic_l0_soln} and taking the real part of the result. The auxiliary quantity \(F_{\alpha\beta\gamma}\) is given by
    \begin{multline}
   F_{\alpha\beta\gamma}(U) = \int \frac{dU}{C_{\alpha\beta\gamma}(U)} \\[1.2ex]= \begin{cases}
            -\recip{\gamma}\log\!\left(\frac{\alpha^2}{\gamma}- U\right) & \text{if} \;\; \beta=0, \\[1ex]
            -\frac{1}{\beta^2 (U-U_0)} & \text{if} \;\; \Delta U = 0,\\[1.7ex]
            \frac{\arg(U-U_0-i\Delta U)}{\beta^2\Delta U} & \text{if} \;\;\Delta U \ne 0, \;\gamma \in I, \\[1.7ex]
            \frac{-i\log\left(\frac{U-U_0+|\Delta U|}{U-U_0-|\Delta U|}\right)}{2\beta^2\Delta U} & \text{ if}\;\; \Delta U \ne 0, \;\gamma \in I^* \setminus  I.
        \end{cases}
    \end{multline} 
    
\begin{figure}[htbp]
    \includegraphics{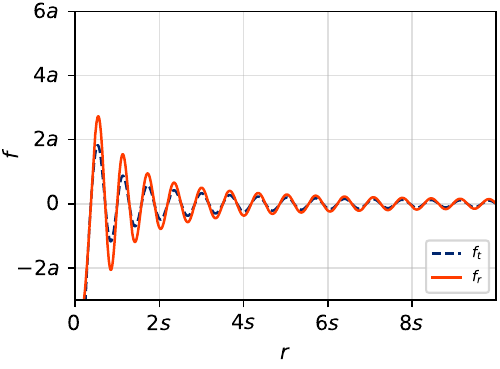}    
    \caption[Monopole perturbations for timelike \(b_\mu\) with \(\beta = \alpha\)]{ \(\hat f_{ t}, \hat f_{ r}\) with \(\beta = \alpha\), \(\gamma = \frac52 \alpha\beta\), \(\omega = 2\pi\), and \(s=1\).}
    \label{fig:ftt}
    \end{figure}

\begin{figure}[htbp]
    \includegraphics{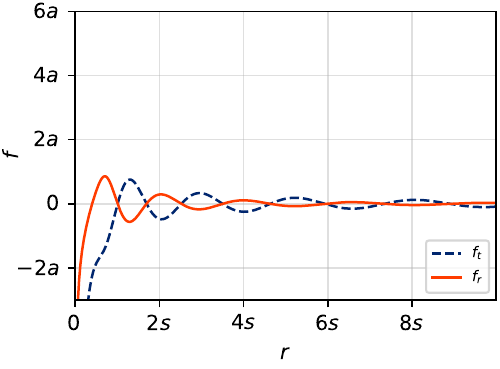}
    \caption[Monopole perturbations for \(\beta = -\alpha\)]{ \(\hat f_{ t}, \hat f_{ r}\) with \(\beta = -\alpha\), \(\gamma = \frac52 \alpha\beta\), \(\omega = 2\pi\), and \(s=1\).}
    \label{fig:fts}
    \end{figure}

\begin{figure}[htbp]
    \includegraphics{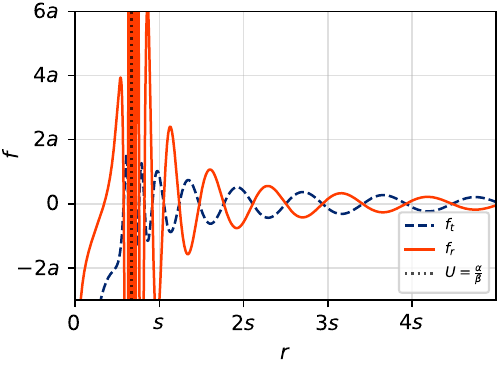}
    \caption[Monopole perturbations for critical spacelike \(b_\mu\)]{ \(\hat f_{ t}, \hat f_{ r}\) with \(\beta = -2\alpha\), \(\gamma \approx 4\alpha\beta\), \(\omega = 2\pi\), and \(s=1\).}
    \label{fig:ftcs}
    \end{figure}

    The single-frequency perturbations obtained by taking \(a(\omega) = a\delta(\omega-\omega_0)\), with \(\omega_0=2\pi\) and \(s=1\), are shown in Figures \ref{fig:ftt}, \ref{fig:fts}, and \ref{fig:ftcs}. The vertical axes are marked in terms of the amplitude coefficient \(a\). As expected, the perturbations become amplified near the critical configurations.
    
    \subsection{Asymptotics for Time-Independent Multipoles}
        For higher-multipole perturbations, we assume time-independence and decompose the perturbations in terms of spin-weighted spherical harmonics \({}_sY_{\ell m}\) \cite{swss:2007, swss:2025} as follows
        \begin{equation}
        \begin{gathered}
            f_t = \sum_{\ell=0}^\infty \sum_{m=-\ell}^\ell f^{\ell m}_t \;{}_0Y_{\ell m}, \qquad f_{r} = \sum_{\ell =0}^\infty
 \sum_{m=-\ell}^{\ell} f_{r}^{\ell m} \;{}_0Y_{\ell m}, \\
 f_{\theta} \pm \frac{i}{\sin\theta}f_{\phi} = \recip2 \sum_{\ell=0}^\infty \sum_{m=-\ell}^{\ell} \left(f^{\ell m}_{\theta} \pm if^{\ell m}_{\phi}\right) {}_{\pm1} Y_{\ell m}.
 \end{gathered}\end{equation}
    The \(\ell =0\) solution corresponds to the time-independent limit of \eqref{eq:quadratic_l0_soln}. 
    
        For \(\ell \ge 1\), the equations of motion can be reduced to two algebraic relations and two decoupled differential equations. The relations take the form
        \begin{equation}
            \begin{gathered}
                f_{ r}^{\ell m } = -\frac{T(r)}{R(r)} f_{ t}^{\ell m }, \\[1ex]
                f_{\phi}^{\ell m} = \frac{i}{2}\ell(\ell + 1) \int \left( \frac{\sqrt{\frac sr}}{U} -\frac{T(r)}{R(r)}\right) f^{\ell m}_{ t}\, dr, 
            \end{gathered} 
            \label{eq:sph_constr}
        \end{equation}
        where the auxiliary quantities \(T(r), R(r)\) are given by
        \begin{equation}
            \begin{aligned}
                T(r) &= \frac{3\ell(\ell+1)}{8\kappa r^2}\sqrt{\frac sr} + b^t b^r, \\
                R(r)&= \frac{3\ell(\ell+1)}{8\kappa s}U\partial _r U+C_{\alpha\beta\gamma}(U).
            \end{aligned}
        \end{equation} 
        
        The remaining differential equations determine the overall behavior of the perturbations. The component \(f_\theta ^{\ell m}\) decouples from all other components and satisfies
        \begin{equation}
            H\!\left(\ell, -\ell-1; -1; \tfrac rs\right) f_\theta^{\ell m}=0,
        \end{equation}
        where \(H\) is the hypergeometric differential operator:
        \begin{equation}
            H(a, b; c;z) u = z(1-z)u'' + (c-(a+b+1)z)u'-abu.
        \end{equation} 
        Solutions of this equation can be constructed in terms of the hypergeometric function \({}_2 F_1(a, b;c;z)\).

        Introducing \(f_t^{\ell m} = \frac{u}{r}\) removes the first-derivative term, yielding 
        \begin{equation}
                u'' + \frac{4\kappa r^2 b_t \left(b^t - b^r \frac{T}{R}\right)-\ell(\ell+1)}{r(r-s)}u = 0. \label{eq:temporal_multipole}
        \end{equation} 
        We derive asymptotic solutions to this equation near the event horizon and infinity.
        
        Far away from the event horizon, neglecting \(O\!\left(\frac{1}{r^3}\right)\), equation \eqref{eq:temporal_multipole} reduces to
        \begin{equation}
        \begin{gathered}
        u''+\frac{\frac14-w^2}{r^2}u=0, \\[1.5ex]
        w = \sqrt{\frac14 -\frac{\ell(\ell+1)}2\left(\frac{(\alpha+\beta)^2  + 2\gamma}{ (\alpha+\beta)^2  - \gamma}\right)}.
        \end{gathered}
        \end{equation}
        The leading-order behavior of the temporal component is therefore
        \begin{equation}
           {
            f_t^{\ell m} \approx \sqrt\frac{s}{r}\left(a^{\ell m} \,r^w + \frac{c^{\ell m}}{r^{w}}\right),
            }
        \end{equation}
        where \(a^{\ell m}, c^{\ell m}\) are constants.

        Near \(r= s\), equation \eqref{eq:temporal_multipole} becomes
        \begin{equation}
            u''+\frac{\ell(\ell+1)}{2s}\left(\frac{1}{r-s}+\frac{3(\gamma-\gamma_\ell)}{s\alpha^2}\right)u=0,
            \label{eq:multipole_horizon}
        \end{equation}
        where \(\gamma _\ell = \frac{3\ell(\ell+1)}{8\kappa s^2}\).

        If \(\gamma \ne \gamma_\ell\), equation \eqref{eq:multipole_horizon} reduces to a Whittaker equation with solution
        \begin{equation}
        {
        \begin{gathered}
        f_t ^{\ell m} \approx \frac sr \left(\tilde\sigma^{\ell m} W_{\nu, \frac 12}(x) + \tilde\rho^{\ell m} M_{\nu, \frac 12}(x) \right), \\[1ex]
        x = \frac{\ell(\ell+1)}{2\nu}\left(\frac rs-1\right), \\[1ex]
        \nu = -i\frac{|\alpha|}4\sqrt\frac{2\ell(\ell+1)}{3(\gamma-\gamma_\ell)},
        \end{gathered}}
        \label{eq:whittaker}
        \end{equation}
        where \(W_{\nu, \mu}, M_{\nu, \mu}\) are the Whittaker functions \cite[\href{https://dlmf.nist.gov/13.14}{(13.14)}]{NIST:DLMF}, and \(\tilde\sigma^{\ell m}, \tilde\rho^{\ell m}\) are constants. Both of these solutions remain finite as we approach the event horizon \(x\to0\).

        The case \(\gamma = \gamma_\ell\) must be treated separately, since the Whittaker parameter \(\nu\) diverges as \(\gamma \to \gamma_\ell\). In this case, equation \eqref{eq:multipole_horizon} reduces to a Bessel equation, with solution
        \begin{equation}
        {
        \begin{gathered}
        f^{\ell m}_t \approx \frac{s}{r}x\left(\sigma^{\ell m} J_1(x) + \rho^{\ell m} Y_1(x)\right),\\
        x = \sqrt{2\ell(\ell+1)\left(\frac rs - 1\right)},
        \end{gathered}}
        \label{eq:bessel}
        \end{equation}
        where \(J_1, Y_1\) are the Bessel functions of the first and second kinds, respectively, and \(\sigma^{\ell m}, \rho^{\ell m}\) are constants. Although \(Y_1(x)\) diverges as \(x\to0\), the combination \(xY_1(x)\) remains finite. Consequently, both independent solutions are regular near the horizon. In the degenerate case \(\gamma = \gamma_\ell\), the leading-order near-horizon behavior is independent of the background parameters \(\alpha, \beta\).
        
        To smoothly recover the Bessel solution in the limit \(\gamma\to \gamma_\ell\), the Whittaker coefficients must scale according to:
        \begin{equation}
            \begin{aligned}
                \tilde\rho^{\ell m} &= 2\nu\,\Gamma(-\nu)\frac{\rho^{\ell m}}{\pi},  \\ 
                \tilde\sigma^{\ell m} &= 2\nu\left[\sigma^{\ell m} + \left(\log \nu - \psi(1-\nu) - \frac{1}{2\nu}\right)\frac{\rho^{\ell m}}{\pi}\right],
            \end{aligned}
        \end{equation}
        where \(\Gamma(z)\) is the gamma function and \(\psi(z)\) is the digamma function. 

    The remaining components \(f_{ r}^{\ell m}\) and \(f_{\phi}^{\ell m}\) can then be obtained from \eqref{eq:sph_constr}.

\section{Conclusion}

    In this work, we investigated equilibrium configurations and linear perturbations of a bumblebee vector field with a symmetry-breaking quartic potential in a fixed Schwarzschild spacetime. 
    
    For the monopole sector, the characteristic quadratic \(C_{\alpha\beta\gamma}(U)\) controls both the existence of the background solution and the behavior of linear perturbations.  
    Requiring the field to be real and regular yields explicit bounds on the background parameters and identifies critical configurations where the characteristic quadratic acquires a real root. Near these configurations, the monopole perturbations become enhanced, suggesting that the linear approximation breaks down and motivating a complete nonlinear analysis in the future.
    
    For higher multipoles, we show that the asymptotic behavior at infinity obeys a power law whose exponents depend on the background parameters. Near the event horizon, the perturbations admit a general solution in terms of Whittaker functions and reduce to Bessel functions for the special values of \(\ell\) satisfying \(\gamma = \gamma_\ell\), when such values exist. In this degenerate case, the leading-order behavior is independent of the background parameters \(\alpha, \beta\).
    
    These results provide an analytical description of small departures from static equilibrium for Lorentz-violating vector fields in black hole spacetimes. Natural extensions of this work include the analysis of time-dependent multipoles, a more detailed study of the behavior near the singular points of the perturbation equations, and an investigation of the gravitational backreaction of the vector field.

\appendix 

\section{Alternative expressions for the background solution}
The raised components of the background solution in Gullstrand-Painlev\'e coordinates are
\begin{equation}
    \begin{aligned}
        b^t &= \frac{\alpha}U+\beta - \frac{\sqrt\frac sr}U \text{sign}(\alpha) \sqrt{C_{\alpha\beta\gamma}(U)}, \\
        b^r &= -\text{sign}(\alpha) \sqrt{C_{\alpha\beta\gamma}(U)}.
    \end{aligned}
    \label{eq:raised_gp}
\end{equation}

The lowered components of the background solution in Schwarzschild coordinates are
\begin{equation}
    \begin{aligned}
        b_t &= \alpha+\beta U, \\
        b_r &= \text{sign}(\alpha)\frac{\sqrt{C_{\alpha\beta\gamma}(U)}}{U}.
    \end{aligned}
\end{equation}

The raised components of the background solution in Schwarzschild coordinates are
\begin{equation}
    \begin{aligned}
        b^t &= \frac{\alpha}{U} + \beta, \\
        b^r &= -\text{sign}(\alpha) \sqrt{C_{\alpha\beta\gamma}(U)}.
    \end{aligned}
\end{equation}

\bibliography{refs}

\end{document}